\documentclass[final,5p,times,twocolumn,authoryear]{elsarticle}

\usepackage{amssymb}
\usepackage{lipsum}
\usepackage{amsmath}
\usepackage[dvipsnames]{xcolor}
\usepackage{ulem}

\begin{document}

\begin{frontmatter}


\title{Quantification of 2D vs 3D BAO tension using SNIa as a redshift interpolator and test of the Etherington relation}

\author[first,second]{Arianna Favale}\ead{afavale@roma2.infn.it}
\author[second]{Adrià Gómez-Valent}\ead{agomezvalent@icc.ub.edu}
\author[first]{Marina Migliaccio}\ead{migliaccio@roma2.infn.it}

\affiliation[first]{organization={Dipartimento di Fisica and INFN Sezione di Roma 2, Università di Roma Tor Vergata},
            addressline={via della Ricerca Scientifica 1}, 
            city={Roma},
            postcode={00133}, 
            country={Italy}}

\affiliation[second]{organization={Departament de Física Quàntica i Astrofísica and Institut de Ciències del Cosmos, Universitat de Barcelona},
            addressline={Av. Diagonal 647}, 
            city={Barcelona},
            postcode={08028}, 
            country={Spain}}

\begin{abstract}
Several studies in the literature have found a disagreement between compressed data on Baryon Acoustic Oscillations (BAO) derived using two distinct methodologies: the two-dimensional (2D, transverse or angular) BAO, which extracts the BAO signal from the analysis of the angular two-point correlation function; and the three-dimensional (3D or anisotropic) BAO, which also exploits the radial clustering signal imprinted on the large-scale structure of the universe. This discrepancy is worrisome, since many of the points contained in these data sets are obtained from the same parent catalogs of tracers and, therefore, we would expect them to be consistent. Since BAO measurements play a pivotal role in the building of the inverse distance ladder, this mismatch impacts the discourse on the Hubble tension and the study of theoretical solutions to the latter. So far, the discrepancy between 2D and 3D BAO has been only pointed out in the context of fitting analyses of cosmological models or parametrizations that, in practice, involve the choice of a concrete calibration of the comoving sound horizon at the baryon-drag epoch. In this Letter, for the first time, we quantify the tension in a much cleaner way, with the aid of apparent magnitudes of supernovae of Type Ia (SNIa) and excluding the radial component of the 3D BAO. We avoid the use of any calibration and cosmological model in the process. At this point we assume that the Etherington (a.k.a distance duality) relation holds. We use state-of-the-art measurements in our analysis, and study how the results change when the angular components of the 3D BAO data from BOSS/eBOSS are substituted by the recent data from DESI Y1. We find the tension to exist at the level of $\sim 2\sigma$ and $\sim 2.5\sigma$, respectively, when the SNIa of the Pantheon+ compilation are used, and at $\sim 4.6\sigma$ when the latter are replaced with those of DES Y5. In view of these results, we then apply a calibrator-independent method to investigate the robustness of the distance duality relation when analyzed not only with 3D BAO measurements, but also with 2D BAO. This is a test of fundamental physics, which covers, among other aspects, variations of the speed of light with the cosmic expansion or possible interactions between the dark and electromagnetic sectors. We do not find any significant hint for a violation of the cosmic distance duality relation in any of the considered data sets. 
\end{abstract}

\begin{keyword}
distance ladder \sep baryon acoustic oscillations \sep cosmological tensions \sep tests of $\Lambda$CDM
\end{keyword}

\end{frontmatter}


\section{Introduction}\label{sec:introduction}

Baryon Acoustic Oscillations (BAO), as standardizable rulers, and Supernovae of Type Ia (SNIa), as standardizable candles, stand out as two pivotal observational tools in cosmology. They were crucial for the discovery \citep{SupernovaSearchTeam:1998fmf,SupernovaCosmologyProject:1998vns} and subsequent characterization of the late-time acceleration of the universe \citep{2dFGRS:2005yhx,SDSS:2005xqv} and, hence, for the consolidation of $\Lambda$CDM as the current standard cosmological model \citep{Peebles:2002gy,Padmanabhan:2002ji,Turner:2022gvw}. BAO and SNIa are a source of precious background information \citep{eBOSS:2020yzd,Brout:2022vxf,Rubin:2023ovl,DES:2024hip,DESI:2024mwx}. They are excellent independent probes of the geometry and energy content of the universe, complementary to Cosmic Microwave Background (CMB) experiments \citep{Aghanim:2018eyx,ACT:2020gnv}, weak gravitational lensing measurements \citep{Kilo-DegreeSurvey:2023gfr} and a plethora of emerging cosmological probes like cosmic chronometers, gamma-ray bursts or quasars, among others \citep{Moresco:2022phi}. 

Importantly, SNIa and BAO are also instrumental in the discussion of the Hubble tension. The mismatch between the direct cosmic distance ladder measurement of $H_0$ by the SH0ES Team \citep{Riess:2021jrx}, which is model-independent, and the value inferred by the {\it Planck} Collaboration from the analysis of the CMB temperature, polarization and lensing spectra assuming $\Lambda$CDM \citep{Aghanim:2018eyx} is already reaching the $5\sigma$ level. When combined with SNIa, anisotropic (or 3D) BAO, calibrated with the value of the comoving sound horizon at the baryon-drag epoch measured by {\it Planck} considering standard physics before the decoupling time ($r_d\sim 147$ Mpc), lead to values of $H_0$ fully in accordance with the Planck/$\Lambda$CDM estimate, rendering the tension with the local measurement high, see e.g.  \citep{Bernal:2016gxb,Feeney:2018mkj,Lemos:2018smw,DES:2024ywx}. This method for estimating $H_0$ using $r_d$ as an anchor is known as the inverse distance ladder (IDL). The only way to reconcile the IDL result obtained with 3D BAO with the SH0ES measurement keeping standard pre-recombination physics is by a sudden growth of the SNIa absolute magnitude accompanied by a fast phantom-like increase of $H(z)$ at $z\lesssim 0.1$ \citep{Alestas:2020zol,Alestas:2021luu,Gomez-Valent:2023uof,Tutusaus:2023cms,Bousis:2024rnb}, while sticking to the Planck/$\Lambda$CDM behavior at higher redshifts. It is well-known that a phantom transition alone cannot solve this problem \citep{Benevento:2020fev,Camarena:2021jlr,Efstathiou:2021ocp,Heisenberg:2022gqk}. Alternatively, a very local  transition in the SNIa absolute magnitude at $z\lesssim 0.01$ \citep{Marra:2021fvf,Perivolaropoulos:2022vql,Perivolaropoulos:2022khd,Ruchika:2023ugh} or the existence of unaccounted for systematic effects in the first or second rungs of the direct distance ladder could also do the job \citep{Desmond:2019ygn,Desmond:2020wep,Wojtak:2023sts,Hogas:2023pjz,Wojtak:2024mgg}.

Paradoxically, when the IDL is built using angular (transversal or 2D) BAO data one gets a very different answer \citep{Camarena:2019rmj}. In this case, one can explain with late-time new physics the large values of the Hubble constant and absolute magnitude of SNIa measured by SH0ES respecting the constancy of the latter \citep{Gomez-Valent:2023uof}. The effective dark energy density must become, though, negative at $z\gtrsim 2$ and, hence, transversal BAO require new physics at much higher redshifts \citep{Gomez-Valent:2023uof}\footnote{Analyses using binning with cosmic chronometers and SNIa also suggest hints of negative dark energy densities at $z > 1$ \citep{Colgain:2022rxy,Malekjani:2023ple}. This preference is observed at approximately $2\sigma$ C.L.}. 
\cite{Akarsu:2023mfb} have found that, in the context of a model with a sign-switching cosmological constant, it is possible to get rid of the $H_0$ and growth tensions when angular BAO is employed in the fitting analysis, instead of 3D BAO (see also \citealt{Anchordoqui:2024gfa}). Fitting results improve even more if dark energy is of quintessence type at late times, after the transition, something that can be realized in the model with phantom matter proposed by \cite{Gomez-Valent:2024tdb}. Interestingly, a negative cosmological constant could also induce a larger abundance of extremely massive galaxies at $z\gtrsim 5$ and, therefore, have a positive bearing on the tension with the data from the {\it James Webb Space Telescope}  \citep{Adil:2023ara,Menci:2024rbq}. Analyses of interacting dark energy \citep{Bernui:2023byc} and bimetric gravity \citep{Dwivedi:2024okk} carried out with 2D and 3D BAO data lead to completely inconsistent results too.

\begin{figure}
    \centering
    \includegraphics[scale=0.64]{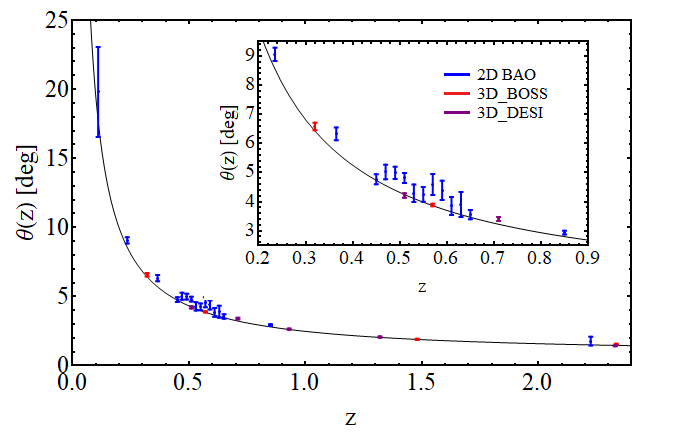}
    \caption{2D and 3D BAO measurements of $\theta(z)=r_d/D_M(z)$, see Tables \ref{tab:BAO_data} and \ref{tab:BAO_data2D}, and Sec. \ref{sec:MethodData}. In black, we show the theoretical curve obtained with the best-fit Planck/$\Lambda$CDM model (TT,TE,EE+lowE+lensing analysis, \citealt{Aghanim:2018eyx}). The inner plot shows a zoom in the redshift range $z\in [0.2,0.9]$.}
    \label{fig:theta(z)}
\end{figure}

The aforementioned conflict between anisotropic and angular BAO data is evident from Fig. \ref{fig:theta(z)}. It clearly points to the existence of unknown systematic errors affecting one or both BAO data sets, or to an underestimation of their uncertainties. 2D BAO analyses make use of the angular two-point correlation function or the angular power spectrum, and measure the angular position of the BAO peak. They do not require the use of a fiducial model to convert redshifts and angles into positions in a 3D tracer map \citep{Sanchez:2010zg}. This is why it is usually claimed to be only weakly dependent on a cosmological model (see e.g. \citealt{deCarvalho:2021azj}). 

Anisotropic BAO, instead, employ a fiducial cosmological model to build the 3D maps of tracers in redshift space. The impact of the fiducial cosmology has been demonstrated to induce only small shifts in the inferred cosmological distances when it does not differ significantly from the true model \citep{Carter:2019ulk,Heinesen:2019phg,Bernal:2020vbb,Pan:2023zgb,Sanz-Wuhl:2024uvi,Chen:2024tfp}. However, \cite{Anselmi:2022exn} argued that these works only check the unbiasedness of the results, not the uncertainties, and do not explore a wide enough range of the fiducial parameter values to be compatible with many of the measured BAO distances, which extend to parameter values far from the Planck/$\Lambda$CDM best-fit model. In fact, \cite{Anselmi:2018vjz} showed that fitting the two-point function while fixing the cosmological and the non-linear-damping parameters at
fiducial values leads to an underestimation of the errors by a factor of two, which could certainly mitigate the tension between the 3D and 2D BAO measurements.

While this discrepancy as well as its impact on the Hubble tension has been pointed out previously in the literature \citep{Camarena:2019rmj,Gomez-Valent:2023uof}, so far there has been no accurate quantification of the statistical tension between these two BAO data sets. In this work, we provide a method to perform this quantification in a quite model-independent way. We apply it to state-of-the-art BAO data, including the anisotropic measurements reported by DESI in their first year release, and compare these results with those obtained using the 3D BAO data from BOSS/eBOSS. Moreover, we also study the effect of a possible underestimation of the 3D BAO uncertainties on our results. This is motivated by the work by \cite{Anselmi:2018vjz}.

In the first part of the Letter we take the fulfillment of the Etherington relation \citep{Etherington:1933}, a.k.a distance duality relation (DDR),  for granted. In terms of the redshift $z$, the latter reads,

\begin{table*}[t!]
    \centering
    \begin{tabular}{cccc}
        \hline
        Survey & $z$ & $D_A/r_d$& References \\
        \hline
        \hline
            BOSS DR12 & 0.32 & $6.5986\pm0.1337$ & \cite{Gil_Mar_n_2016}\\
         & $0.57$  & $9.389\pm0.103$ & \\
         \hline
         eBOSS DR16Q & 1.48  & $ 12.18\pm 0.32$ & \cite{Hou:2020rse}\\
           \hline
         Ly$\alpha$-F eBOSS DR16 & $2.334$   & $11.25^{+0.36}_{-0.33}$ & \cite{duMasdesBourboux:2020pck}\\
         \hline\hline
         LRG1  DESI Y1 & $0.51$   & $9.02\pm 0.17$ & \cite{DESI:2024uvr}\\
         \hline
         LRG2  DESI Y1 & $0.71$   & $9.85\pm 0.19$ & \cite{DESI:2024uvr}\\
         \hline
         LRG3+ELG1  DESI Y1 & $0.93$   & $11.25\pm 0.16$ & \cite{DESI:2024uvr}\\
         \hline
         ELG2  DESI Y1 & $1.32$   & $11.98\pm 0.30$ & \cite{DESI:2024uvr}\\
         \hline
         Ly$\alpha$-F DESI Y1 & $2.33$   & $11.92\pm 0.29$ & \cite{DESI:2024lzq}
        \\
        \hline
    \end{tabular} \caption{List with the anisotropic BAO data points used in this work. See the quoted references for details. As explained in Sec. \ref{sec:MethodData}, we use two alternative 3D BAO data sets, one containing the BOSS/eBOSS data points and another one with the DESI Y1 data. 
    }\label{tab:BAO_data}
\end{table*}

\begin{equation}
D_L(z) = (1+z)^2 D_A(z)\,,
\end{equation}
with $D_L(z)$ and $D_A(z)$ the luminosity and angular diameter distances, respectively.
The Etherington relation holds in metric theories of gravity with photons conserved and propagating in null geodesics. Thus, any deviation from this relation, which would imply that 

\begin{equation}\label{eq:etaDef}
\eta(z)\equiv \frac{D_L(z)}{(1+z)^2 D_A(z)}\ne 1
\end{equation}
in some redshift range(s), would hint at new physics either in the gravity sector or beyond the standard model of particle physics. Among other possibilities, this could be the case of models with a time-varying speed of light (see e.g. \citealt{Lee:2020zts,Gupta:2023mgg}) or theories with a coupling between photons and fields in the dark sector, e.g. with axions or axion-like particles (see e.g. \cite{Jaeckel:2010ni,Carosi:2013rla}). Of course, these deviations could also hint at the existence of unknown systematic errors in the data. 

Tests of the Etherington relation have been carried out using SNIa in combination with 3D BAO \citep{More:2008uq,Nair:2012dc,Ma:2016bjt,EUCLID:2020syl,Renzi:2021xii}, cosmic chronometers \citep{Avgoustidis:2010ju}, strong lensing time delays \citep{Qi:2024acx} or data on compact radio quasars \citep{Tonghua:2023hdz}, and even with luminosity distances from high-redshift quasars and data from strong gravitational lenses \citep{Qin:2021jqy}. See also \citep{Qi:2019spg}. They scrutinize a broader range of assumptions compared to other more specific tests, such as those conducted to study the validity of the cosmological principle, see, e.g., \citep{Bengaly:2021wgc,Favale:2023lnp}. No significant deviation from the DDR has been found so far. In view of the tension between the 2D and 3D BAO data sets, we deem interesting and very timely to also study what happens if we employ angular BAO, instead of 3D BAO, to test the Etherington relation and see whether the tension between these two data sets reflects also into differences at the level of the DDR. Moreover, we will compare the results obtained with anisotropic BAO from BOSS/eBOSS and DESI Y1 to determine if conclusions on the validity of the Etherington relation have changed with the new data. It is of utmost importance to check whether this relation holds, since it is taken for granted in our quantification of the tension between the 2D and 3D BAO. Thus, any significant departure from the DDR would invalidate the results obtained in the first part of the paper.

\begin{table*}
    \centering
    \begin{tabular}{ccccc}
        \hline
       Survey & $z$ & $\theta_{BAO}$ [deg] & References \\
        \hline
        \hline
        SSDS DR12 & 0.11 & $19.8 \pm3.26$ & \cite{deCarvalho:2021azj} \\ \hline
        SDSS DR7 & 0.235 & $9.06 \pm 0.23$ & \cite{Alcaniz:2016ryy} \\
        &0.365 & $6.33 \pm 0.22$ \\ \hline
        SDSS DR10 &0.45 & $4.77\pm0.17$ & \cite{Carvalho:2015ica} \\
        &0.47 & $5.02 \pm0.25$ \\
        &0.49 & $4.99 \pm0.21$ \\
        &0.51 & $4.81 \pm 0.17$ \\
       & 0.53 & $4.29 \pm 0.30$ \\
        &0.55 & $4.25 \pm 0.25$ \\ \hline
        SDSS DR11 &0.57 & $4.59 \pm 0.36$ & \cite{Carvalho:2017tuu} \\
        &0.59 & $4.39 \pm 0.33$ \\
        &0.61 & $3.85 \pm 0.31$ \\
        &0.63 & $3.90 \pm 0.43$ \\
        &0.65 & $3.55 \pm 0.16$ \\ \hline        
       DES Y6  & 0.85 & $2.932\pm 0.068$ & \cite{DES:2024cme}\\ \hline
       BOSS DR12Q  &2.225 & $1.77 \pm 0.31$ & \cite{deCarvalho:2017xye}     
        \\
        \hline
    \end{tabular} \caption{List with the 16 2D BAO data points used in this work, with $\theta_{\rm BAO}(z)\,[{\rm rad}]=r_d/[(1+z)D_A(z)]$. The 2D BAO data point of DES Y6 has been computed using the values of $\alpha=(D_A/r_d)/(D_A/r_d)|_{\rm Planck}$ extracted with the angular correlation function (ACF) and the angular power spectrum (APS) methods, which read $\alpha_{\rm ACF}=0.952\pm 0.023$ and $\alpha_{\rm APS}=0.962\pm 0.022$, respectively, with a Pearson correlation coefficient $\rho=0.863$ \citep{DES:2024cme}. We have considered the full covariance matrix in the weighted average computation of the central value and the uncertainty. See the quoted references for further details.}\label{tab:BAO_data2D}
\end{table*}

This Letter is structured as follows. In Sec. \ref{sec:MethodData} we explain the methodology and the data sets employed in our analyses. We describe the method used to quantify the tension between the angular and anisotropic BAO measurements and also the calibrator-independent test of the Etherington relation. In Sec. \ref{sec:results} we present our results and discuss how they change when we substitute the BOSS/eBOSS data by the recent DESI data, and also the impact of the SNIa sample employed in the analysis. Finally, in Sec. \ref{sec:conclusions} we present our conclusions and outlook. 


\section{Methodology and data}\label{sec:MethodData}

 We use the comoving sound horizon $r_d$ as a cosmic standard ruler. Galaxy surveys have been able to measure the angle  

\begin{equation}\label{eq:sr}
\theta(z) = \frac{r_d}{D_M(z)}
 \end{equation}
 at various redshifts, with $D_M(z)=(1+z)D_A(z)$ the comoving angular diameter distance. We employ the angular and anisotropic BAO data listed in Tables \ref{tab:BAO_data} and \ref{tab:BAO_data2D}, respectively. In the case of the 3D BAO, we use two alternative data sets: one contains the data from BOSS/eBOSS, whereas the other contains the data from DESI Y1. We refer to these two data sets as 3D\_BOSS and 3D\_DESI, respectively. The BOSS/eBOSS and DESI compressed data points have been obtained from catalogs with partially overlapping population and non-disjoint patches in the sky, so we opt not to combine them and proceed as explained above in order to avoid double counting issues, since the correlations between these two sets of data are not known.

The model- and calibrator-independent methods that we will employ to quantify the tension between the 3D and 2D BAO data sets and the test of the DDR make only use of the angular component of the 3D BAO measurements. Thus, we exclude the radial components and the data points provided in terms of the dilation scale.

As already explained in the Introduction, our first aim in this Letter is to quantify the tension between the angular and anisotropic BAO data sets. It is clear that this quantification cannot be carried out through a direct (one-to-one) comparison of the data points contained in the two BAO data sets, since their redshifts are different. In order to do it in a model-independent and parametrization-independent way, we make use of SNIa and the method described below.

Let us start with a brief description of the SNIa data. The relation between the luminosity distance and the apparent magnitude $m$ of a standard candle is given by

\begin{equation}\label{eq:sc}
m(z) = M+25+5\log_{10}\left(\frac{D_L(z)}{\rm 1\, Mpc}\right)\,,
\end{equation}
where $M$ is its absolute magnitude. In this paper, we employ SNIa as standard candles. We make use of two different SNIa samples separately: the  Pantheon+ compilation \citep{Scolnic:2021amr}\footnote{https://github.com/PantheonPlusSH0ES/DataRelease}, which is a collection of 1701 lightcurves of 1550 distinct SNIa in the redshift range $0.001<z<2.26$. In particular, we consider a subsample of 1624 data points, since we do not include those SNIa lying in Cepheid host galaxies, although we have explicitly checked that the impact of these objects on our results is completely negligible; and the recently released DES Y5 compilation \citep{DES:2024tys}\footnote{https://github.com/des-science/DES-SN5YR}, which consists of a total of 1829 SNIa spanning the redshift range $0.025<z<1.13$. For cosmological purposes, the fundamental difference between these two SNIa samples is the higher statistics of DES Y5 compared to Pantheon+, especially for $z>0.5$, where the number of high-quality SNIa is fivefold those of Pantheon+. This difference will be particularly relevant in our discussion since most of the BAO data points lie in that region.
 
 From Eqs.  \eqref{eq:sr} and \eqref{eq:sc} it is straightforward to write the luminosity and angular diameter distances in terms of the calibrators ($M$ and $r_d$) and the measured quantities ($m(z)$ and $\theta(z)$). Plugging these expressions into Eq. \eqref{eq:etaDef} we obtain, 

 \begin{equation}\label{eq:etacal}
\eta(z)=\frac{10^{m(z)/5}\theta(z)}{(1+z)}\frac{10^{-5-M/5}{\rm Mpc}}{r_d}\,.
\end{equation}
If the Etherington relation is fulfilled (i.e., if $\eta(z)=1$) and if we have measurements of $\theta(z)$ and $m(z)$ at the same redshift, we can compute the degeneracy direction in the calibrators' plane,  

\begin{equation}\label{eq:degcal}
\bar{r}_d10^{M/5} = \frac{10^{m(z)/5}\theta(z)}{10^5(1+z)}\,,
\end{equation}
with $\bar{r}_d\equiv r_d/(1\,{\rm Mpc})$ the dimensionless sound horizon. This calculation is independent of the curvature of the universe, which we do not need to specify. We do not have SNIa and BAO data at the very same redshifts, of course, but this is not a major complication, since we can employ some interpolation method to obtain the constrained values of $m(z)$ at the BAO redshifts.


One of these methods is the Gaussian Processes (GP) regression technique \citep{2006gpml.book.....R}, which can be used to reconstruct functions from Gaussian distributed data under very minimal assumptions. Thanks to the definition of a kernel function, which in most cases depends only on two hyperparameters, one can track the correlations between points where data are absent. In principle, one should marginalize over these hyperparameters to account for the propagation of their uncertainties to the reconstructed function. However, due to the large covariance matrix of the SNIa samples, this process is very expensive from a computational point of view. In this work, we opt to follow the same approach already tested in \citep{Favale:2023lnp}, where it is shown that differences in the reconstructed shapes obtained with the marginalization and optimization procedures are smaller than $0.1\sigma$ in the case under study, thus not impacting significantly the final results. We make use of the publicly available \texttt{GaPP} code developed by \cite{Seikel:2012uu}.
Using GP we generate $N=10^{4}$ samples of SNIa apparent magnitudes at the BAO redshifts $z_i$. For each redshift $z_i$, we also draw $N$ realizations of $\theta(z_i)$ from a Gaussian distribution that has as mean the measured value and as standard deviation the associated uncertainty. With this information we can obtain a
distribution of \textit{N} samples of
the product $\bar{r}_d10^{M/5}$ (Eq. \ref{eq:degcal}) at the BAO redshifts. There exist non-null correlations, since we employ correlated SNIa data. On the other hand, we expect non-zero positive correlations between several data points in the 2D and 3D data samples, simply because they have been obtained from the same parent catalogs of tracers (cf. Tables \ref{tab:BAO_data} and \ref{tab:BAO_data2D}). However, these correlation coefficients have not been quantified, so we cannot use them in our analysis. This fact will lead us to estimate, in practice, a conservative lower bound of the tension between the angular and anisotropic BAO data sets, see Sec. \ref{sec:tension} for details.

\begin{figure*}[t!]
    \centering
    \includegraphics[scale=0.095]{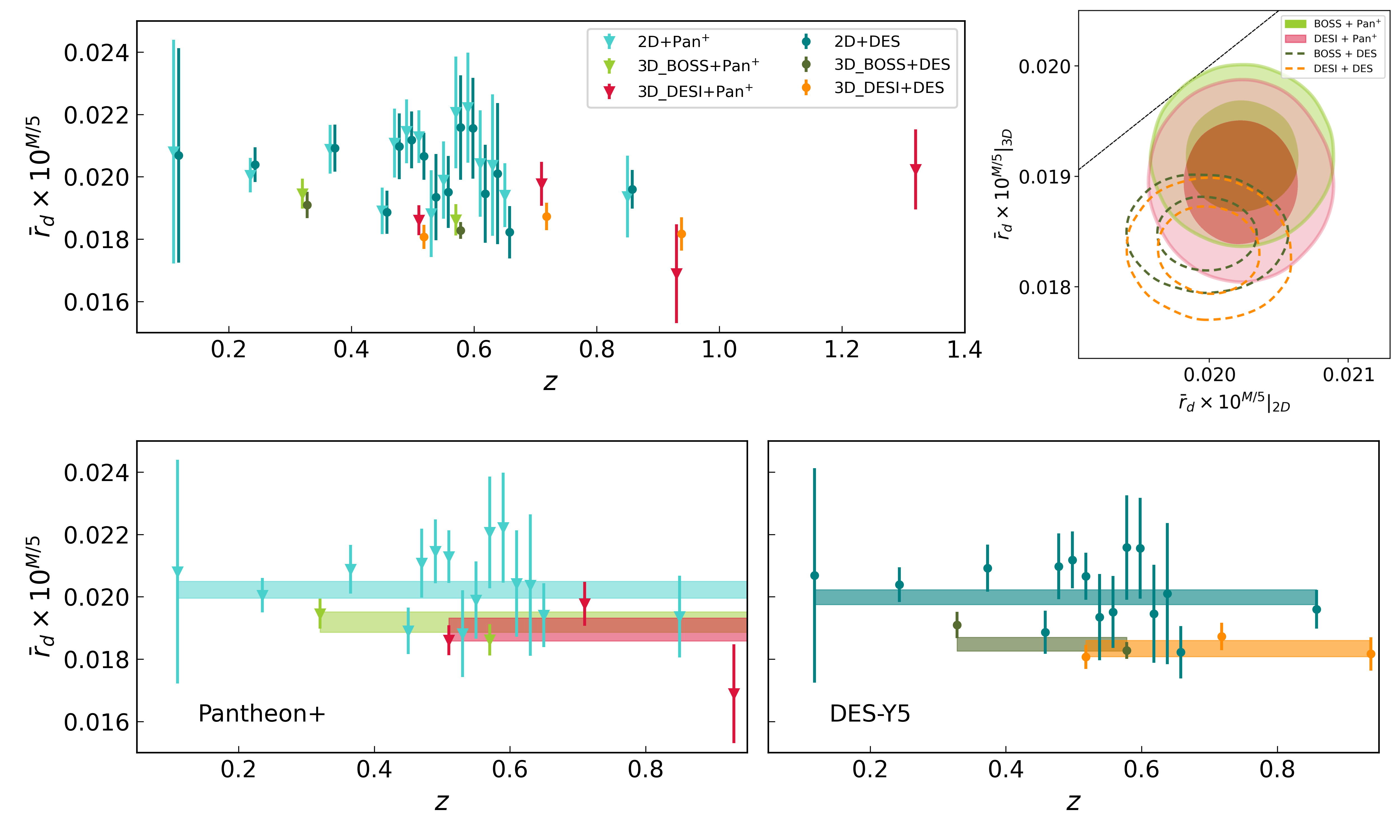}
    \caption{{\it Left upper plot:} Constraints at $68\%$ C.L. on the product $\bar{r}_d\times 10^{M/5}$ obtained making use of the SNIa of Pantheon+ or DES Y5 and the individual 3D and 2D BAO data points. We follow the method explained in Sec. \ref{sec:MethodData}. We plot the points obtained with DES Y5 shifted by $\Delta z = 0.01$ to visually distinguish them from those obtained with Pantheon+. We only show the points at 
 $z< 1.4$ because at higher redshifts error bars are large and exceed the displayed range of values. {\it Lower plots:} Same as the left upper plot, but with a zoom in the range $z\in [0.05,0.95]$ using the Pantheon+ and DES Y5 samples, respectively. The horizontal bands represent the 2D, 3D\_BOSS and 3D\_DESI BAO constraints at $68\%$ C.L. {\it Right upper plot}: The corresponding contours at 68\% and 95\% C.L obtained using \texttt{GetDist} \citep{Lewis:2019xzd}. The dashed straight line in black represents the null hypothesis of consistency between the 3D and 2D BAO data sets, i.e.,  $\bar{r}_d\cdot 10^{M/5}|_{\rm 3D}=\bar{r}_d\cdot 10^{M/5}|_{\rm 2D}$.}
    \label{fig:tens2D3D}
\end{figure*}

We can measure the same quantity $\bar{r}_d\times 10^{M/5}$ from the BAO data at different redshifts, which is very convenient since it allows us to establish a bridge between the various BAO data and use it to quantify the tension. The next step is to estimate representative values of $\bar{r}_d10^{M/5}$ for the 2D and 3D BAO data sets. To do so we need to obtain first their distribution out of the Markov chain. This will allow us to quantify the tension between the two data sets, by means of the tension between the corresponding values of $\bar{r}_d10^{M/5}$. As a starting point, we employ the so-called Edgeworth expansion (see \cite{Amendola:1996xwd} and references therein) to compute an analytical approximation of the underlying (exact) distribution,

\begin{equation}\label{eq:Edgeworth}
f(\vec{x})= G(\vec{x},\lambda)\left[1+\frac{1}{6}k^{ijk}h_{ijk}(\vec{x},\lambda)+...\right]\,,
\end{equation}
 which is a multivariate function with the same dimensionality as the total number of BAO data points contained in the 3D and 2D data sets. This expression is valid when the departures from Gaussianity are sufficiently small. The second term inside the brackets and the dots account for non-Gaussian corrections. We have explicitly checked that in the case under study it is more than sufficient to keep only the first correction.  In Eq. \eqref{eq:Edgeworth},  $x^i=d^i-\mu^i$, with $d^i=\bar{r}_d10^{M/5}|_i$ and $\vec{\mu}$ the mean vector, i.e. $\mu^i=\langle\bar{r}_d10^{M/5}|_i\rangle$. The subscript $i$ labels the BAO data point, which can belong to the 2D or 3D BAO data sets. $\lambda=C^{-1}$ is the inverse of the covariance matrix, with elements $C^{ij}=\langle x^i x^j\rangle$. $G(\vec{x},\lambda)$ is the multivariate Gaussian distribution built from that mean and covariance matrix, and $k^{ijk}=\langle x^ix^jx^k\rangle$ are the elements of the so-called skewness matrix. Finally,

\begin{equation}\label{eq:HermiteT}
h_{ijk}(\vec{x})=\lambda_{in}\lambda_{jt}\lambda_{kl}x^nx^tx^l-(\lambda_{ij}\lambda_{kt}+\lambda_{ik}\lambda_{jt}+\lambda_{jk}\lambda_{it})x^t
\end{equation}
is the Hermite tensor of order 3. In Eqs. \eqref{eq:Edgeworth} and \eqref{eq:HermiteT} we have employed  Einstein's summation convention. All these objects can be directly computed from the chain. Once we build the distribution of Eq. \eqref{eq:Edgeworth}, we can sample it treating it as a two-dimensional distribution for $\bar{r}_d10^{M/5}$, with one dimension for the 2D BAO data set, and the other for the 3D BAO data set, i.e. $ f(x_{\rm 2D},x_{\rm 3D})$. When the non-Gaussian features are negligible, the resulting distribution reduces of course to a Gaussian with the following weighted mean vector $\tilde{\mu}$ and inverse covariance matrix $\tilde{\lambda}$,

\begin{equation}\label{eq:Gapprox}
\tilde{\mu}_I =\tilde{\lambda}^{-1}_{IJ}v_J \qquad ;\qquad \tilde{\lambda}_{IJ} = \sum_{i\in I}\sum_{j\in J}\lambda_{ij}\,,
\end{equation} 
with 

\begin{equation}\label{eq:Gapprox2}
v_I = \sum_{i\in I}\sum_{j=1}^{N}\lambda_{ij}\mu_j
\end{equation}
and $N$ the total number of BAO data points contained in the 2D and 3D data sets. Here we have indicated sums over all the elements contained in the data set $I$ as $\sum_{i\in I}$, with $I$ referring to the 2D or 3D BAO data sets. If we perform the analysis with 3D\_BOSS and Pantheon+, $N=20$; if, instead, we use 3D\_DESI, $N=21$. When we employ DES Y5, our analysis is restricted to $z<1.13$, since this is the maximum redshift of the SNIa sample. Hence, our BAO data set reduces to $N=17$ if we employ 3D\_BOSS and $N=18$ if we employ 3D\_DESI.
 
To quantify the tension between the anisotropic and angular BAO data we can just build the histogram of $\chi^2 = -2\ln f(x_{\rm 2D},x_{\rm 3D})$ resulting from the sampling of the two-dimensional distribution obtained with the procedure described above, and then compute the $p$-value associated to the hypothesis of having only one single BAO data set (built out from the joint 2D and 3D BAO data sets). In point of fact, we can obtain the full distribution of $p$-values, which allows us to compute also the associated uncertainty. The smaller the $p$-value, the more the hypothesis is excluded, of course, indicating greater inconsistency between the two data sets. We repeat this exercise with both 3D\_BOSS and 3D\_DESI (one at a time) in combination with 2D BAO to determine if the level of tension between the angular and the anisotropic BAO data sets changes. In addition, we also test the impact of the SNIa sample by replacing Pantheon+ with DES Y5. We present the results of this analysis in Sec. \ref{sec:tension}.

In the second part of the paper, we test the Etherington relation following the calibrator-independent method employed by \cite{Tonghua:2023hdz}. It is based on the quantity

\begin{equation}\label{eq:etapairs}
\eta_{i,j} \equiv \frac{\eta(z_i)}{\eta(z_j)}=\frac{10^{m(z_i)/5}\theta(z_i)}{10^{m(z_j)/5}\theta(z_j)}\left(\frac{1+z_j}{1+z_i}\right)\,,
\end{equation}
which is built from the ratio of Eq. \eqref{eq:etacal} at two different redshifts. Avoiding the use of the calibrators of the distance ladders in this test is important not to propagate possible biases that might have been introduced in their measurement. This is relevant on account of the Hubble tension. Departures of $\eta_{i,j}$ from unity could hint at a violation of the DDR. In Sec. \ref{sec:EtherigtonTest} we show our results obtained with 2D, 3D\_BOSS and 3D\_DESI BAO, in combination with the SNIa of the Pantheon+ or DES Y5 compilations.


\section{Results}\label{sec:results}

\subsection{Tension between angular and anisotropic BAO}\label{sec:tension}

In the left upper plot of Fig. \ref{fig:tens2D3D} we show the constraints on the product $\bar{r}_d\times 10^{M/5}$ defined in Eq. \eqref{eq:degcal}, obtained by making use of the individual BAO angles in combination with the SNIa apparent magnitudes of Pantheon+ or DES Y5. In the lower plots we also present the overall constraints obtained from each BAO data set, see the horizontal bands at 68\% C.L. and the results reported in Table \ref{tab:results}. The contour plot on the top right complements the preceding information and helps to better assess visually the tension between the 2D and 3D BAO data sets. The latter points to a $\sim 2\sigma$ C.L. tension when the anisotropic BOSS data are used, and grows up to the $\sim 2.5\sigma$ C.L. with DESI. This result is obtained employing SNIa from Pantheon+. If, instead, we perform the analysis using the DES Y5 SNIa sample, the statistical significance of the tension reaches the $\sim 4.6\sigma$ level with both BOSS and DESI. This increase is mainly driven by the high number of SNIa observed within the redshift range $0.5\lesssim z \lesssim 0.9$ by DES Y5, which is roughly five times the number of SNIa in the same range as reported by Pantheon+ \citep{DES:2024tys}. Most of the BAO data used in this work lie precisely in that region (see, respectively, the left upper plot and the horizontal bands in the lower panels of Fig. \ref{fig:tens2D3D})\footnote{The replacement of the SNIa from Pantheon+ with those from DES Y5 also enhances the evidence for dynamical dark energy from $2.5\sigma$ to  $\sim 4\sigma$ C.L. when these data sets are combined with DESI+CMB data \citep{DESI:2024mwx}. Although the discussion at hand is different from the one in that paper, this is another manifestation of the significant statistical power of the DES Y5 SNIa compilation. Given the impact that the SNIa data have in these analyses and the differences observed with Pantheon+ and DES Y5, it will be important to track the effect of systematic errors in these samples \citep{Efstathiou:2024xcq}.}.

\begin{table*}[t!]
\centering
\begin{tabular}{ccccc}          
\hline\noalign{\smallskip}        
  BAO data set & $\bar{r}_d10^{M/5}$ & $p$-value & $\bar{r}_d10^{M/5}$ & $p$-value  \\
  \hline\noalign{\smallskip} 
  & \multicolumn{2}{c}{Pantheon+} & \multicolumn{2}{c}{DES Y5} \\
  \hline\hline\noalign{\smallskip}  
   2D  & $(20.23\pm0.27)\cdot10^{-3}$ & $-$ & $(19.98\pm0.24)\cdot10^{-3}$ & - \\
 \noalign{\smallskip} \hline\noalign{\smallskip}
    3D\_BOSS  & $(19.19\pm0.33)\cdot10^{-3}$ & $0.048_{-0.009}^{+0.008}$ & $(18.48\pm0.22)\cdot10^{-3}$ & $<10^{-5}$
 \\
\noalign{\smallskip}    
      3D\_BOSS$^{\star}$& $(19.10\pm0.48)\cdot10^{-3}$ & $0.116_{-0.008}^{+0.016}$ & $(18.47\pm0.38)\cdot10^{-3}$ & $<10^{-3}$ \\
      \noalign{\smallskip}
\hline\noalign{\smallskip}
     3D\_DESI & $(18.95\pm0.37)\cdot10^{-3}$ & $0.018_{-0.002}^{+0.011}$ & $(18.34\pm0.26)\cdot10^{-5}$ & $<10^{-5}$ \\
    \noalign{\smallskip}
      3D\_DESI$^{\star}$ & $(18.98\pm0.53)\cdot10^{-3}$ & $0.105_{-0.008}^{+0.014}$ & $(18.29\pm0.43)\cdot10^{-3}$ & $<10^{-3}$
   \\
   \noalign{\smallskip}\hline
\end{tabular}\caption{Constraints at 68\% C.L. on the product $\bar{r}_d10^{M/5}$ for the 2D and 3D BAO data sets obtained using the SNIa from Pantheon+ and DES Y5, following the procedure explained in Sec. \ref{sec:MethodData}. They incorporate possible deviations from Gaussianity, encapsulated in Eq. \eqref{eq:Edgeworth}, which are, in any case, derisory, as we have explicitly checked by comparing these results with those obtained with Eqs. \eqref{eq:Gapprox}-\eqref{eq:Gapprox2}. The level of tension between the 2D and 3D BAO data sets is quantified by the $p$-value, reported in the third and fifth columns of the table for the case of Pantheon+ and DES Y5, respectively. In the second rows of 3D\_BOSS and 3D\_DESI, we present the results obtained by doubling the uncertainties associated with the 3D BAO measurements. These results are labeled with an asterisk.}\label{tab:results}
\end{table*}

It is also important to note that the roughly circular shape of the contour lines in Fig. \ref{fig:tens2D3D} responds to the fact that the correlations introduced by SNIa turn out to be very small and we have not considered correlations between the 2D and 3D BAO data sets (see Sec. \ref{sec:MethodData}). A positive correlation between these two data sets would leave the individual uncertainties of $\bar{r}_d\times 10^{M/5}$ stable, but the shape of the contour lines would be more elongated in the direction of the straight line shown in the top right plot of Fig. 2. Therefore, the effective distance (measured in terms of $\Delta\chi^2$) between the best-fit point and the straight line would be larger and the tension between the 3D and 2D BAO data would increase. This is why we actually obtain a lower bound of the tension by neglecting the existing correlations between the anisotropic and angular BAO data.

Following the methodology explained in Sec. \ref{sec:MethodData}, we also compute the distribution of $p$-values associated with the hypothesis of having only one single BAO data set. We obtain a $p$-value of $0.048_{-0.008}^{+0.009}$ and $0.018_{-0.002}^{+0.011}$ when employing the SNIa of Pantheon+ in combination with 3D\_BOSS and 3D\_DESI, respectively. When we use the SNIa of DES Y5, instead, the $p$-values become smaller than $10^{-5
}$, see Table \ref{tab:results}. These more quantitative results are fully aligned with those of Fig. \ref{fig:tens2D3D} and highlight a significant tension between the two types of BAO data. This tension persists across different SNIa compilations but is particularly pronounced with the DES Y5 data.

It is crucial to emphasize again that this significant tension is derived in a largely model-independent way, assuming only the validity of the Etherington relation. The addition of the radial component of the 3D BAO data, which has not been considered in our analysis because it is not feasible to incorporate it following a model- and calibrator-independent approach, increases the tension even more. This explains why fitting analyses with angular and anisotropic BAO data lead to completely different results, as shown in, for example, \cite{Bernui:2023byc}. In particular, they require different solutions to the Hubble tension \citep{Gomez-Valent:2023uof}.

Several recent works have pointed out that the DESI BAO data obtained from luminous red galaxies at $z=0.51$ are possibly driving the signs of dynamical dark energy found by combining DESI with CMB, observational Hubble data and SNIa \citep{DESI:2024mwx,Colgain:2024xqj,Carloni:2024zpl, Wang:2024rjd}. It is therefore worthwhile to closely examine these data and their impact on our results. However, the $\sim2\sigma$ offset of the DESI data at $z=0.51$ from the Planck/$\Lambda$CDM model reported by \cite{DESI:2024uvr} is only found in the radial measurement, $D_{H}(z)/r_d = c/H(z)r_d$, i.e., it is not present in the transversal BAO component (see their Figure 15 and also our Fig. \ref{fig:theta(z)}). The latter is fully consistent with Planck/$\Lambda$CDM, as are the other 3D BAO data. Thus, we can conclude that this data point is not playing a major role in our analysis, apart from tightening the constraint on the product $\bar{r}_d\times 10^{M/5}$ extracted with 3D\_DESI.

Finally, motivated by the work by \cite{Anselmi:2018vjz}, we have also studied to what extent the tension loosens if we increase the uncertainties of the 3D BAO measurements by a factor of two. The results are displayed in Table \ref{tab:results}. The $p$-values are approximately 0.1 in this case when the 2D BAO is combined with both 3D BAO data sets using the SNIa of Pantheon+, resulting in a discrepancy of less than $2\sigma$ and, hence, in no significant statistical tension. When the SNIa of Pantheon+ are replaced with those of DES Y5 the statistical tension remains, instead, at $\sim 3.5\sigma$ C.L. An underestimation of the 3D BAO uncertainties as the one estimated by \cite{Anselmi:2018vjz} could mitigate the tension between angular and anisotropic measurements, improving in this way their concordance. However, the tension persists at high confidence level in the light of the most complete SNIa sample even when the uncertainties of the anisotropic BAO data are increased by a factor two.


\subsection{Test of the Etherington relation}\label{sec:EtherigtonTest}

\begin{figure*}[t!]
    \centering
    \includegraphics[scale=0.08]{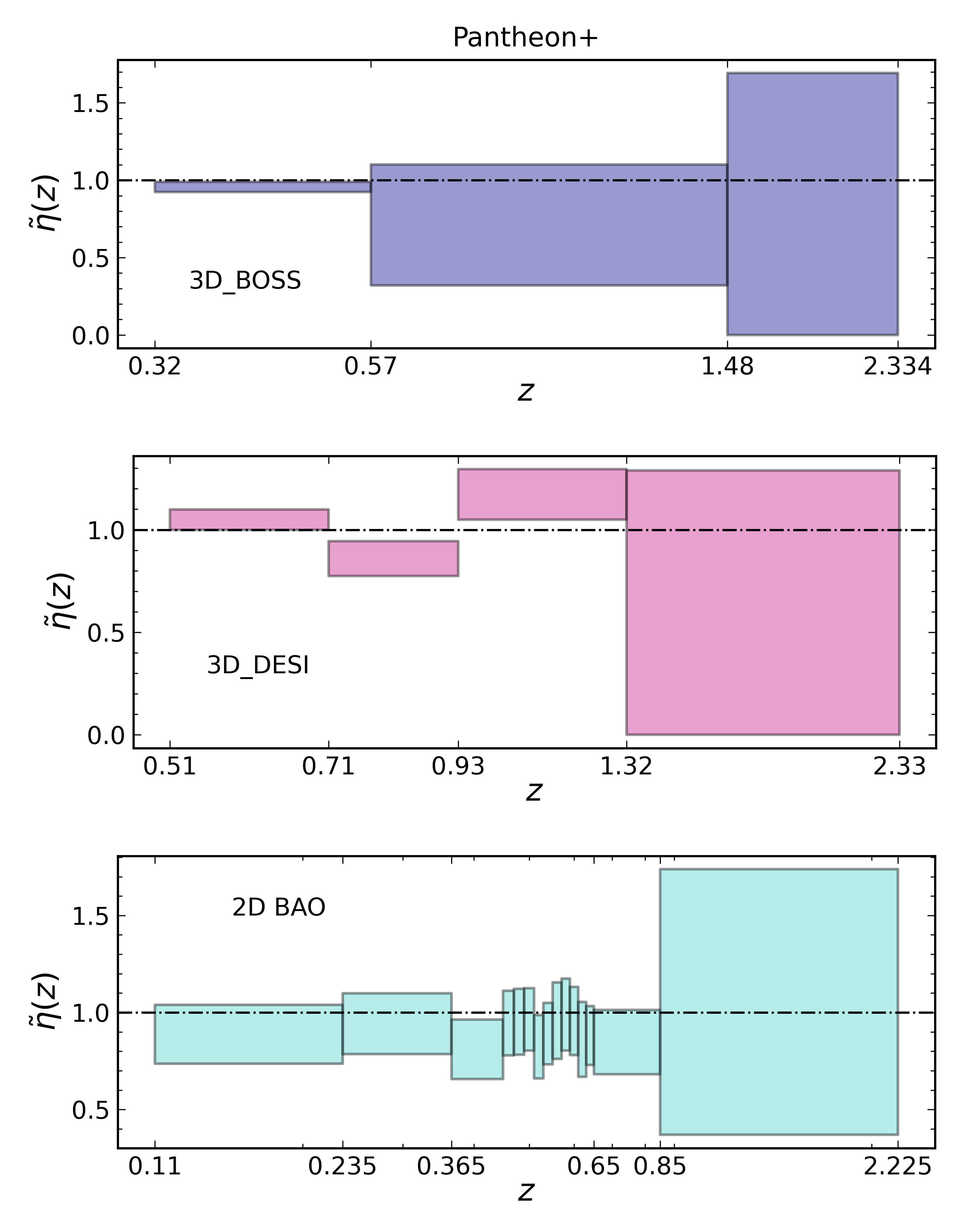}
    \includegraphics[scale=0.08]{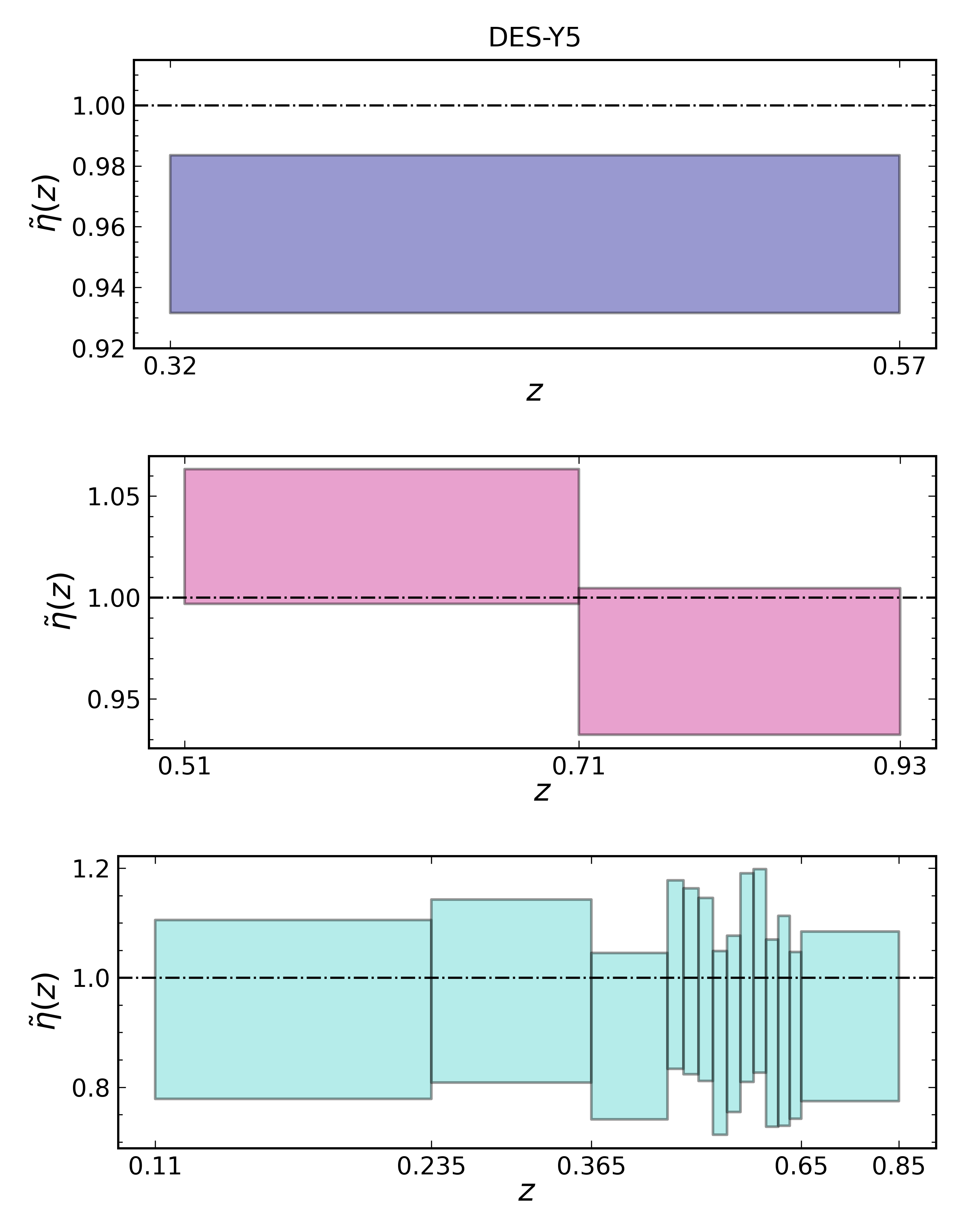}

    \caption{Constraints at $68\%$ C.L. on $\tilde{\eta}(z)$, as defined in Eq. \eqref{eq:etaz}, obtained making use of the 3D and 2D BAO data sets in combination with the SNIa of the Pantheon+ and DES Y5 compilations. The $x$-axes are in logarithmic scale for the sake of a better visualization. We write the $z$ values of the various BAO data points employed in the construction of $\tilde{\eta}(z)$, except in the region of redshifts in the 2D BAO plot where the density of points is too high, see Table \ref{tab:BAO_data2D}.}
    \label{fig:Etherington2D}
\end{figure*}

We follow the procedure described in the last part of Sec. \ref{sec:MethodData} to obtain the values of $\eta_{i,i+1}$ (Eq. \ref{eq:etapairs}) using the various BAO and SNIa data sets. For each of them, we compute the following function 

\begin{equation}\label{eq:etaz}
\tilde{\eta}_I(z) =\sum_{i=1}^{N_I-1}\eta_{i,i+1}\Theta(z-z_i)\Theta(z_{i+1}-z)\,,
\end{equation}
where $\Theta(\cdot)$ is the Heaviside step function, and we show the resulting $68\%$ C.L. constraints on it in Fig. \ref{fig:Etherington2D}. Eq. \eqref{eq:etaz} accounts for all the independent $N_I-1$ couples of $z$. The subscript $I$ labels the BAO+SNIa data set employed in the calculation and $N_I$ is the number of BAO data points contained in it. Eq. \eqref{eq:etaz} allows us to present the results in a very visual way and maximizes the statistical content of our sample without unnecessary repetition. There are only $N_I-1$ independent values of $\eta_{i,j}$ out of the $N_I(N_I-1)$ non-zero values that can be computed using the $N_I$ values of $\eta(z_i)$, with $i\in I$.

Eq. \eqref{eq:etaz} can be used to detect deviations from the DDR. In each of the bins, the quantity $\eta_{i,j}\ne 1$ if $\eta(z_i)\ne\eta(z_j)$. This can only happen if $\eta(z_i)\ne 1$ or/and $\eta(z_j)\ne 1$. This obviously implies a violation of the DDR. On the other hand, $\eta_{i,j}=1$ if $\eta(z_i)=\eta(z_j)$. If the derivative of $\eta(z)$, $f(z)\equiv d\eta/dz,$ is well-defined, we can always do

\begin{equation}
   \eta(z) = \eta(z=0)+\int_0^{z}f(z^\prime) dz^\prime \,.
\end{equation}
Assuming that $z_j>z_i$, the condition $\eta(z_i)=\eta(z_j)$ is satisfied if $f(z)=0$ $\forall z\in [z_i,z_j]$. Hence, our diagnostic is sensitive to sources of violation of the DDR in the redshift interval between $z_i$ and $z_j$. Thus, our estimator $\tilde{\eta}_I(z)$ is able to detect violations of the Etherington relation by construction. The only limitation of our method is the precision of the available data.

We do not find any significant hint of a violation of the DDR. This result supports the validity of the Etherington relation, which is used to quantify the tension between the two BAO data sets in Sec. \ref{sec:tension}.  Regardless of the BAO and SNIa data sets employed in the analysis, the quantity $\tilde{\eta}(z)$ is compatible with $1$ at approximately $1-2\sigma$ C.L. While this result is consistent with previous analyses using the anisotropic 3D\_BOSS data set (see e.g., \citealt{Renzi:2021xii}), our study represents the first investigation in the literature to test the DDR using both the 3D and 2D BAO data sets, and the SNIa compilation from DES Y5. The tension between the anisotropic and angular BAO data sets discussed in Section \ref{sec:tension} does not manifest as a deviation from the DDR when considering the angular BAO data. To date, uncertainties on $\tilde{\eta}_I(z)$ are still large, especially at high redshift, where they can reach 100$\%$ due to large errors in SNIa data.


\section{Conclusions and outlook}\label{sec:conclusions}

We have devoted this Letter to the quantification of the existing tension between the angular and anisotropic data on baryon acoustic oscillations, employing a novel and quite model-independent approach which only relies on the fulfilment of the Etherington relation and is also independent of the calibrators of the direct and inverse distance ladders. By construction, our method excludes the radial 3D BAO data. We find a tension at $\sim 2\sigma$ C.L. between the 2D BAO and the anisotropic data set constructed with BOSS/eBOSS data when using the SNIa from the Pantheon+ compilation. This tension increases to approximately 
$\sim 2.5\sigma$ C.L. when we employ the first data release of DESI instead of BOSS/eBOSS. The replacement of the Pantheon+ SNIa with those of DES Y5 triggers a substantial increase of the tension, which now reaches the $4.6\sigma$ C.L. This can be considered a lower bound of the actual tension because we cannot account for the positive correlations between the 2D and 3D BAO data sets, as these correlations have not been reported in the literature.

The statistical significance of the aforesaid discrepancy would be softened if, for some reason, the BAO uncertainties had been underestimated. This is a plausible possibility according to \cite{Anselmi:2018vjz}, who found that the errors in standard anisotropic BAO analyses might be underestimated by a factor of two. In order to assess the impact of this possible underestimation on our results we have repeated the analysis doubling the uncertainties of 3D\_BOSS and 3D\_DESI and find that the tension decreases below the $2\sigma$ C.L. with Pantheon+ and remains at $3.5\sigma$ C.L. with DES Y5. This tells us that the tension between angular and anisotropic BAO can be mitigated if the error bars in the compressed BAO data are not being properly estimated in current analyses. However, it is still pretty high according to the latest SNIa sample.

In view of the existing tension between 2D and 3D BAO, which grows even more in the context of model-dependent analyses thanks to the inclusion of the radial 3D component, we have also deemed interesting to check whether this mismatch can be translated into a deviation from the Etherington relation. This is also a consistency test, since we have assumed the validity of the latter in the first part of the paper. We have applied a calibrator- and model-independent approach previously used by \cite{Tonghua:2023hdz} that lets us obtain constraints on the violation of the distance-duality relation. This is the first time this method is applied using BAO data, and also the first time that the robustness of the Etherington relation is studied using angular BAO and the SNIa of DES Y5. Our constraints are quite loose. They allow for violations of $\sim 20\%$ below $z\sim 0.8-1$ at $68\%$ C.L., and even greater at larger redshifts. When we combine 3D BAO and DES Y5 SNIa, though, the maximum deviations are of about 10\%. We do not find any compelling evidence for departures from the standard scenario neither in the anisotropic nor the 2D BAO data sets. The maximum incompatibility with the DDR is found to exist at  $\sim 1\sigma$ C.L. in some redshift ranges for the 3D\_DESI and 2D BAO data sets, which is not significant at all. Stronger constraints have been obtained in the literature, but they typically require the use of parametrizations of $\eta(z)$ and, hence, additional assumptions to those considered in this Letter.

In the era of precision cosmology and the existing tensions afflicting the standard $\Lambda$CDM model, it is crucial to elucidate what is causing the discrepancies between the various BAO data sets. This is pivotal to find out, for instance, the origin of the Hubble tension \citep{Camarena:2019rmj,Gomez-Valent:2023uof}. It will be important to re-assess the level of tension between angular and anisotropic BAO once the new data releases from DESI and future data from Euclid become available. Hopefully, these surveys will provide both 3D and 2D BAO data, extracted by the same collaborations. This would be extremely healthy and useful for the community in order to test both methods and pinpoint possible sources of biases. In this Letter we have presented a model- and calibrator-independent methodology that can be applied in these future works.


\section*{Acknowledgements}
AF and MM acknowledge support from the INFN project “InDark”. AGV is funded by “la Caixa” Foundation (ID 100010434) and the European Union's Horizon 2020 research and innovation programme under the Marie Sklodowska-Curie grant agreement No 847648, with fellowship code LCF/BQ/PI21/11830027. He is grateful to the Institute of Cosmology and Gravitation of the University of Portsmouth for its kind hospitality during the writing of this paper. MM is also supported by the ASI/LiteBIRD grant n. 2020-9-HH.0 and by the Italian Research Center on High Performance Computing Big Data and Quantum
Computing (ICSC), project funded by European Union - NextGenerationEU - and National
Recovery and Resilience Plan (NRRP) - Mission 4 Component 2 within the activities of Spoke 3
(Astrophysics and Cosmos Observations). AF and AGV acknowledge the participation in the COST Action CA21136 “Addressing observational tensions in cosmology with systematics and fundamental physics” (CosmoVerse).


\bibliographystyle{elsarticle-harv} 
\bibliography{main}

\end{document}